\documentclass[aps,prb,twocolumn,graphicx,superscriptaddress]{revtex4-1}

\usepackage{graphicx}
\begin{document}

\title{Optical evidence of surface state suppression in Bi based topological insulators}

\author{Anjan A. Reijnders}
\email[]{reijnder@physics.utoronto.ca}
\author{Y. Tian}
\author{L. J. Sandilands} 
\author{G. Pohl}
\author{I. D. Kivlichan}
\author{S. Y. Frank Zhao}
\affiliation{Department of Physics \& Institute for Optical Sciences, University of Toronto, 60 St. George Street, Toronto, ON M5S 1A7, Canada}

\author{S. Jia}
\author{M. E. Charles}
\author{R.J. Cava}
\affiliation{Department of Chemistry, Princeton University, Princeton, NJ 08544, USA}

\author{Nasser Alidoust}
\author{Suyang Xu}
\author{Madhab Neupane}
\author{M. Zahid Hasan}
\affiliation {Joseph Henry Laboratory, Department of Physics, Princeton University, Princeton, NJ 08544, USA}

\author{X. Wang}
\author{S. W. Cheong}
\affiliation{Rutgers Center for Emergent Materials and Department of Physics and Astronomy, Rutgers University, 136 Frelinghuysen Road, Piscataway, NJ 08854, USA}

\author{K. S. Burch}
\email[]{ks.burch@bc.edu}
\affiliation{Department of Physics, Boston College, 140 Commonwealth Ave, Chestnut Hill, MA 02467, USA}

\date{\today}

\begin{abstract}
A key challenge in condensed matter research is the optimization of topological insulator (TI) compounds for the study and future application of their unique surface states. Truly insulating bulk states would allow the exploitation of predicted surface state properties, such as protection from backscattering, dissipationless spin-polarized currents, and the emergence of novel particles. Towards this end, major progress was recently made with the introduction of highly resistive Bi$_2$Te$_2$Se, in which surface state conductance and quantum oscillations are observed at low temperatures. Nevertheless, an unresolved and pivotal question remains: while room temperature ARPES studies reveal clear evidence of TI surface states, their observation in transport experiments is limited to low temperatures. A better understanding of this surface state suppression at elevated temperatures is of fundamental interest, and crucial for pushing the boundary of device applications towards room-temperature operation. In this work, we simultaneously measure TI bulk and surface states via temperature dependent optical spectroscopy, in conjunction with transport and ARPES measurements. We find evidence of coherent surface state transport at low temperatures, and propose that phonon mediated coupling between bulk and surface states suppresses surface conductance as temperature rises.
\end{abstract}

\maketitle

\section{Introduction}
Three dimensional topological insulators have recently been at the forefront of condensed matter research due to their novel surface states with linear Dirac dispersion. Theory predicts that while the bulk in TI crystals is insulating, its surface states are conductive and are protected from backscattering, resulting in dissipationless spin-polarized currents.\cite{Qi:2011wt,Moore:2007,Hasan:2010ub} A better understanding of these surface states could lead to novel spintronic devices; faster, smaller, and more efficient transistors; and provide a quantum mechanical environment conducive to fundamental research on Majorana fermions and magnetic monopoles.\cite{Hasan:2010ub,Qi:2011wt} The existence of TI surface states was initially demonstrated by ARPES and STM experiments. However, accessing surface states with more general probes, such as transport and optics, has proven to be exceedingly difficult.\cite{LaForge:2010dx,Ren:2010ji,2011arXiv1101.1315X,Luo:2011vf,2011PhRvB..84w5206J,Xiong:2011tm,2012arXiv1209.3593A,2012PhRvL.108h7403V,Tang:2012wd,Jenkins:2012wh,Sushkov:2010dd,2012PhRvL.108h7403V, Post:2013td} Consequently, several important TI properties remain poorly understood, among which are the coupling of surface to bulk states, electron-phonon coupling, and room temperature surface conductance. A recent breakthrough in TI research was the observation of TI surface state conductance at low temperatures in transport experiments.\cite{Ren:2010ji,2011arXiv1101.1315X,2010NanoL.10.5032S,Kim:2012cs} At room temperature, however, surface conductance is strongly suppressed. This is rather surprising as ARPES experiments still clearly resolve the surface states at room temperature.\cite{Hasan:2010ub,Qi:2011wt} Thus, a pivotal question in the field remains:  Why is coherent surface state conductance in TIs clearly visible at low temperatures, but suppressed in room temperature transport measurements?

By combining ARPES, transport, and optical spectroscopy results, simultaneously probing the bulk and surface states of TIs, and independently extracting their optical conductivities, we shed new light on electron phonon interactions in TI surface states. In particular, temperature dependent optical spectroscopy proves to be an excellent method for elucidating the transport suppression mechanism and surface-to-bulk coupling in TIs, as information about carrier scattering rates, effective mass, optical transitions, phonons, and their respective coupling can be retrieved.

In this article, we explore the optical properties of four Bi-based TIs, Bi$_2$Te$_2$Se, BiSbSe$_2$Te, Bi$_2$Te$_3$, and Bi$_2$Se$_3$, and focus on the temperature dependent reflectance $R(\omega)$ and optical conductivity $\sigma(\omega)$ of Bi$_2$Te$_2$Se over the range of 2.5 meV - 5.95 eV. In this compound we find a low temperature upturn in R($\omega$) in addition to a reflectance increase in the longitudinal optical phonon mode, indicative of metallic behavior. A unique fit, congruent DC resistivity, a matching sign change in the Hall coefficient, and comparison with ARPES data, all on the same crystal, provide strong evidence for the surface state nature of this low temperature conductive channel. Hence, we find Bi$_2$Te$_2$Se to be the most promising candidate for studying surface state suppression.

\begin{figure}
\includegraphics[scale=0.24]{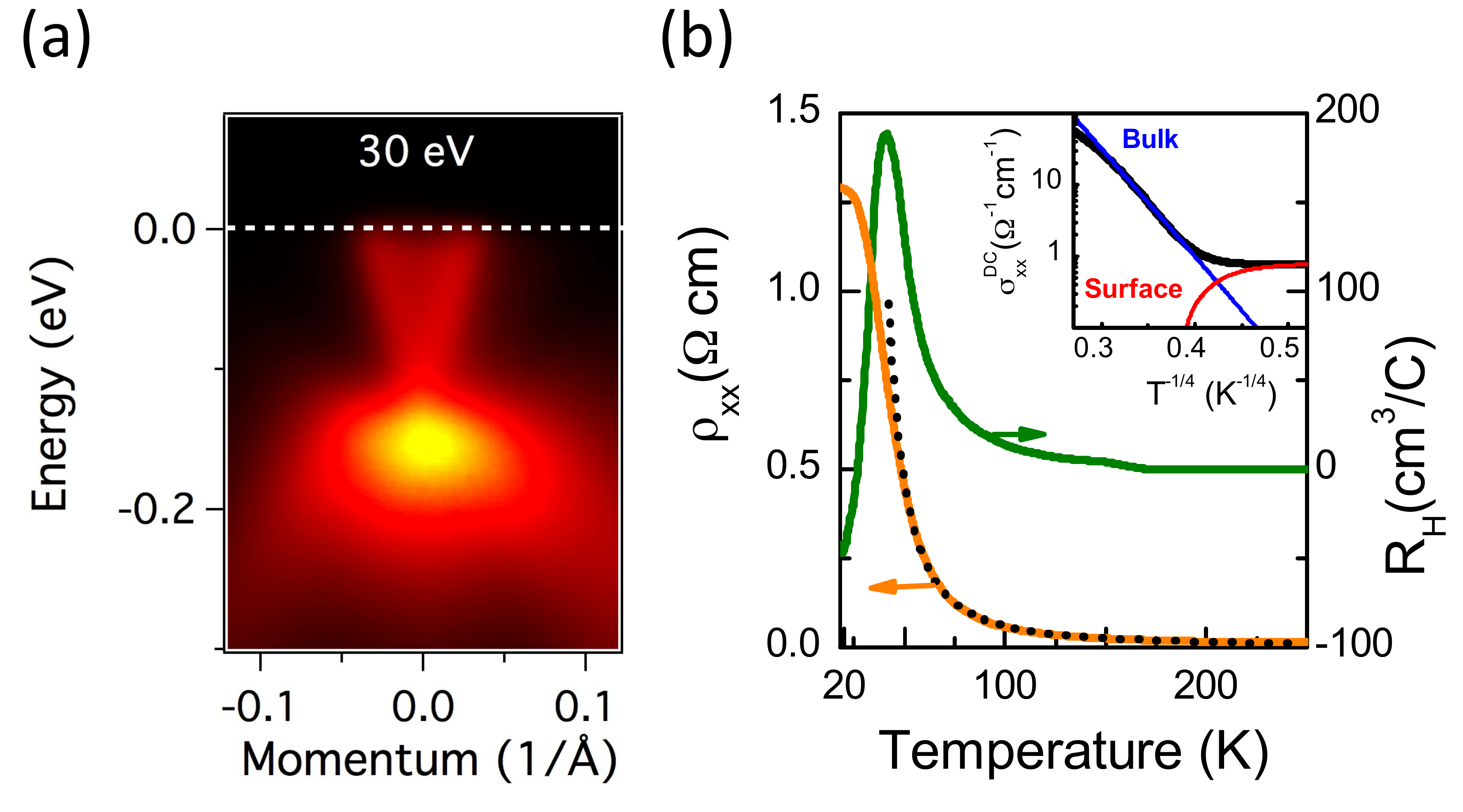}
\caption{\label{fig:BTSFig1}(a) ARPES spectrum of Bi$_2$Te$_2$Se shows the Fermi level lies inside the bulk gap, 100 meV above the Dirac point. (b) Longitudinal resistivity (orange) of Bi$_2$Te$_2$Se reaches 1.3 $\Omega$ $cm$ below 20 K, and the Hall coefficient (green) flips sign below 40 K, indicative of a change in dominant carriers from holes to electrons. An Arrhenius fit (black dotted line) reveals an activation energy of 16 meV. The inset of (b) shows metallic surface state conductance below 43 K (red), after subtracting a 3D variable-range hopping fit (blue) from the DC conductivity of Bi$_2$Te$_2$Se. The Hall coefficient was obtained at magnetic fields of 1.5 T and -1.5 T, and then symmetrized.
}
\end{figure}

Early research on 3D TIs has mostly focused on binary chalcogenides, such as Bi$_2$Se$_3$ and Bi$_2$Te$_3$, in which the Fermi level typically crosses the bulk conduction or valence band, respectively. This results in bulk conductance concealing electronic signatures of the surface states.\cite{Ren:2010ji,2011arXiv1101.1315X,Hasan:2010ub,Qi:2011wt,LaForge:2010dx,Hor:2009hl,Post:2013td} Hence, a natural starting point for studying TIs with the Fermi level inside the band gap is the ternary compound Bi$_2$Te$_2$Se. In this material, Se vacancies are suppressed due to preferential bonding of Se to Bi in the Te-Bi-Se-Bi-Te quintuple layer structure.\cite{Ren:2010ji,2011arXiv1101.1315X} Indeed, our ARPES measurement (Fig.~\ref{fig:BTSFig1}a) shows that the Fermi level of the Bi$_2$Te$_2$Se crystal used in this study lies in the gap, while DC transport reveals a large bulk resistance of 1.3 $\Omega$ cm below 20 K (Fig.~\ref{fig:BTSFig1}b), in good agreement with previous results.\cite{2012PhRvB..85w5406N,Ren:2010ji,2011PhRvB..84w5206J,2011arXiv1101.1315X,Xiong:2011tm} As originally proposed by Z. Ren, {\it et al.},\cite{Ren:2010ji} a closer look at the DC conductivity of Bi$_2$Te$_2$Se reveals a distinct metallic surface contribution, after the subtraction of bulk Variable Range Hopping conductivity (inset of Fig.~\ref{fig:BTSFig1}b). At elevated temperatures, however, surface state conductance appears to diminish and a concurrent sign change in the Hall coefficient (Fig.~\ref{fig:BTSFig1}b) indicates a shift in dominant carrier mobility from electrons to holes.\cite{Ren:2010ji,2011arXiv1101.1315X} Other experiments on Bi$_2$Se$_3$ nano devices,\cite{2010NanoL.10.5032S} and thin exfoliated Bi$_2$Se crystals,\cite{Kim:2012cs} in which the Fermi-level was tuned into the bulk gap, also show strongly temperature dependent surface conductance.
This behavior is consistent with our optics results that reveal coherent surface state transport only below 43K. We expound this suppression by a careful comparison of the temperature dependence of the bulk and surface optical conductivities. Specifically, we provide direct evidence for phonon-mediated suppression of the surface kinetic energy. In addition, our results provide a natural explanation for the sign change of the Hall signal (Fig.~\ref{fig:BTSFig1}b), and can be used as a guideline for the synthesis of novel TIs in which electron-phonon coupling is suppressed.

\section{Experiment}
\subsection{Samples}

Measurements on Bi$_2$Te$_3$, Bi$_2$Te$_2$Se, and BiSbSe$_2$Te were performed on single crystalline samples grown at Princeton University. Sample growth details are described in previously published work.\cite{2011PhRvB..84w5206J} Bi$_2$Se$_3$ single crystals were grown at Rutgers University, using a slow-cooling method with mixtures of high-purity Bi 99.999\%, and Se 99.999\% chemicals. The mixtures were slowly heated in evacuated quartz tubes to 950$^{\circ}$C, where they were held for 10 hours. Afterward, the crystal growth was achieved by cooling the quartz tubes from 950$^{\circ}$C to 200$^{\circ}$C at a rate of 5$^{\circ}$C per hour, followed by furnace cooling. 

Reflectance of all samples was measured in the natural cleavage plane (111), except for a mechanically polished side surface of Bi$_2$Se$_3$. Prior to each reflectance measurement the crystals were cleaved using adhesive tape, to produce a clean and lustrous surface, and were rapidly loaded into a vacuum chamber, limiting the maximum atmospheric exposure to 20 minutes.

\subsection{Experimental setup and measurements}
All reflectance experiments were performed in near-normal geometry over the far through near-infrared range (2.5 meV - 1.24 eV), using a modified Bruker VERTEX 80v FTIR Spectrometer (see Appendix for a detailed description and characterization of the setup). The sample and reference (gold-coated Si wafer) were mounted on a gold-coated cone and attached to a custom made sample holder to prevent detection of unwanted systemic radiation. After laser-alignment of the sample and reference, the sample holder was placed in a high vacuum chamber and moved into the beam path using a Thermionics Z-275 linear translator. In-situ gold coating of the samples was used to produce an ideal reference for each measurement.\cite{1993ApOpt..32.2976H} Between successive experiments of different energy ranges, the evaporated gold was removed using adhesive tape, also cleaving a fresh surface of the sample. Reflectance in each range was measured twice per sample, always reproducing the same spectra. 
An ARS LT-3B continuous flow cryostat was used to cool the sample and reference from 292K to 11 K with 10 intermediate stops. To supplement the spectral range of measured reflectance, a Woollam Inc VB-400 VASE ellipsometer was used at room temperature between 0.75-5.95 eV, for a combined spectral range of 2.5 meV - 5.95 eV. We note that our setup provides excellent reproducibility between thermal cycles (better than 0.4\%), which is pivotal for the observation of subtle changes in temperature dependent reflectance (See Appendix for additional details on the experimental setup).

Transport measurements were performed via standard lock-in techniques using a 10 kOhm ballast resistor and a 2.5 V excitation voltage modulated at 973 Hz. Contacts were applied by hand to the sample in a Hall bar geometry using silver paint. Our measurements showed some contribution from the longitudinal resistance to the transverse signal and vice versa. To correct for this, the data were symmetrized with respect to applied magnetic field (-1.5 T and 1.5 T).

Angle-resolved photoemission spectroscopy (ARPES) measurements of Bi$_2$Te$_2$Se and BiSbSe$_2$Te were performed with incident photon energy of 30 eV and 9 eV, respectively, at the PGM beamline in the Synchrotron Radiation Center (SRC) in Wisconsin. Samples were cleaved \textit{in situ} between $6$ to $20$ K at chamber pressure better than $5 {\times} 10^{-11}$ torr at the SRC, resulting in shiny surfaces. Energy resolution was better than 15 meV and momentum resolution was better than 1\% of the surface  Brillouin Zone. The measurement temperature was $15$ K. A single Dirac cone surface state is observed in both ARPES measurements, demonstrating the topological insulator state in the Bi$_2$Te$_2$Se and BiSbSe$_2$Te crystals. In Bi$_2$Te$_2$Se, the chemical potential is about 100 meV above the Dirac point, which is found to be within the band gap. The Fermi momentum ($k_F$) and the Fermi velocity ($v_F$) of the sample are found to be 0.03 $\AA^{-1}$ and $6  {\times} 10^{5}$ m/s, respectively, consistent with previous studies.\cite{Luo:2011vf,Ren:2010ji,Taskin:2011uq,2011arXiv1101.1315X,2012PhRvB..85w5406N}

\begin{figure}
\includegraphics[scale=0.24]{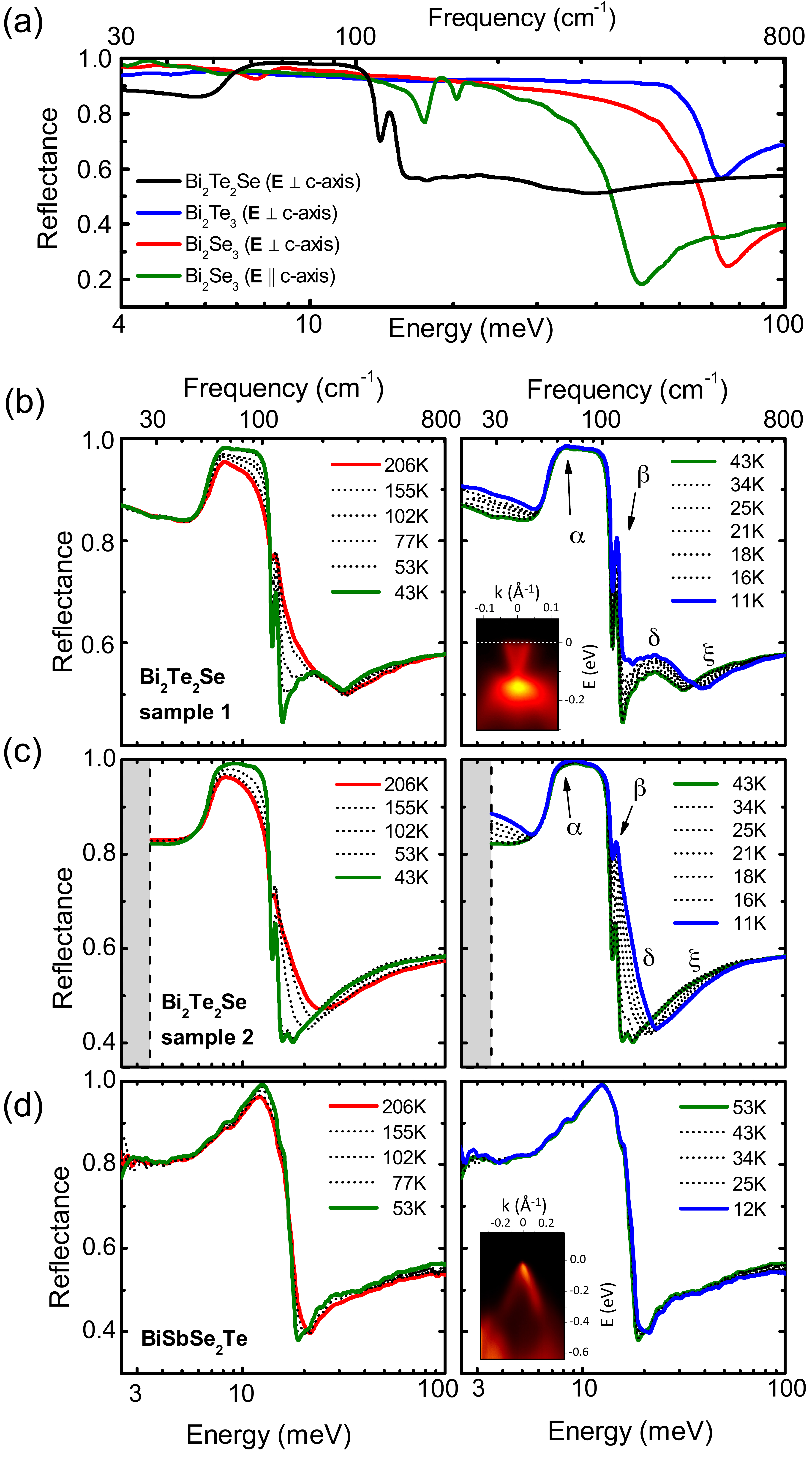}
\caption{\label{fig:BTSFig2}(a) Low temperature infrared reflectance of topological insulators Bi$_2$Se$_3$ (with $\vec{E}$ $\parallel$ and $\perp$ to the c-axis), Bi$_2$Te$_3$, and Bi$_2$Te$_2$Se. The plasma edge reveals the metallic nature of the binary compounds. Far infrared reflectance and ARPES (insert) spectra of Bi$_2$Te$_2$Se samples 1 and 2 (b,c), and BiSbSe$_2$Te (d) are shown at high (left) and low (right) temperature to emphasize their temperature evolution. In particular, we note the upturn in R($\omega$) at low energy below 43 K in Bi$_2$Te$_2$Se}
\end{figure}

\section{Results and Analysis}

\subsection{Reflectance spectra}
The low temperature far infrared (FIR) reflectance spectra of Bi$_2$Te$_2$Se, Bi$_2$Te$_3$, and Bi$_2$Se$_3$ are shown in Fig.~\ref{fig:BTSFig2}a. For an intuitive interpretation of these data the classical Drude model is a useful guide, in which optical conductivity is defined as $\hat{\sigma}(\omega)=\omega_{p}^2/(\gamma-i\omega)$.\cite{Liu:1993vu,YuCard} Here, $\omega_{p}$ is the plasma frequency of the free carriers, and $\gamma=1/\tau$ is the inverse optical lifetime, which is distinct from the quasiparticle lifetime (inverse of the imaginary part of the self energy function) measured by ARPES, and the phase coherence time measured by quantum oscillations.\cite{1985PhRvB..32.8442D,1997CaJPh..75..509M} The plasma frequency, resulting from the collective resonance of free electron oscillations in a 3D parabolic band, is given by $\omega_{p}^{2}=ne^{2}/\epsilon_{0}m_b$, where $n$ is the free carrier density, and $m_b$ is the band mass. Since carriers in a real material are (partially) screened, it's only possible to measure the screened plasma frequency, $\widetilde{\omega}_p =\omega_p/\epsilon_{\infty}$, where $\epsilon_{\infty}$ is the high frequency dielectric constant. This free carrier resonance against a static charged background effectively creates an oscillating dipole that re-emits absorbed radiation for frequencies $\omega<\widetilde{\omega}_p$. Hence, reflectance is nearly 100\% in this regime, with a (local) minimum at $\widetilde{\omega}_p$, indicative of a sample's free carrier density or oscillator strength (in the case of phonons), while the sharpness of the feature (i.e. its slope) is a measure of $\gamma$. The plasma frequency is also a useful measure of the free carrier spectral weight, as $\omega_{p}^{2}\propto\int\sigma_{1}d\omega$. However, it should be noted that in a purely quantum mechanical picture the total spectral weight is truly a measure of the kinetic energy of the carriers, and cannot be simply connected to $n$ or $m_b$.\cite{2007PhRvB..75p5407G} 

A clear example of the plasma frequency is shown in Fig.~\ref{fig:BTSFig2}a. Both Bi$_2$Te$_3$ and Bi$_2$Se$_3$, measured with $\vec{E}$ $\perp$ c-axis (111), have a minimum around 75 meV, and a corresponding large carrier density around 2.4$\times$10$^{20}$ cm$^{-3}$.\cite{Stordeur:1992wq} Since $\omega_p^2$ is inversely proportional to $m_b$ (when interactions are ignored), one would expect a reduced plasma edge in the reflectance of Bi$_2$Se$_3$ measured parallel to the c-axis ($m_{b,\parallel c}/m_{b,\perp c}\approx 4$).\cite{Stordeur:1992wq} This prediction is confirmed in Fig.~\ref{fig:BTSFig2}a, however, reflectance close to 100\% below $\omega_p^{bulk}$ still conceals possible signatures of surface state transport due to their much smaller contribution to reflectance. 
\

In Bi$_2$Te$_2$Se and BiSbSe$_2$Te (Fig.~\ref{fig:BTSFig2}b-d) reflectance is quite different as expected for TIs with the Fermi energy in the bulk gap (see Fig.~\ref{fig:BTSFig3}a for Reflectance data over the full measured range of Bi$_2$Te$_2$Se sample 1). Specifically, the reflectance of Bi$_2$Te$_2$Se contains a peak followed by a plateau and a depression; the quintessential signature of an optical phonon. Similar in appearance to a free carrier plasma edge, the depression corresponds to the phonon's longitudinal optical mode. However, since optical phonons occur at finite frequency, their signature in R$(\omega)$ is preceded by reduced reflectance below its transverse optical mode ($\hbar\omega <$ 7 meV for Bi$_2$Te$_2$Se). This distinguishes the reflectance feature from a free carrier plasma edge, and confirms the reduced carrier density and bulk conductivity compared to Bi$_2$Te$_3$ and Bi$_2$Se$_3$.\cite{Ren:2010ji,2011arXiv1101.1315X,Xiong:2011tm,2011PhRvB..84w5206J,2012PhRvB..85w5406N} This makes Bi$_2$Te$_2$Se a prime candidate for studying the temperature dependence of surface state transport. Indeed, Fig.~\ref{fig:BTSFig2}b,c show that as temperature is reduced below 43 K, an upturn in reflectance appears below 7 meV, while the reflectance minimum at 16 meV increases. This is a signature of an increasing free carrier density in the classical Drude model (since $\omega^2_p\propto n$), which quantum mechanically is associated with a change in the carrier's kinetic energy.\cite{2007PhRvB..75p5407G} As we show later, this can only be attributed to surface states, and is corroborated by our DC conductivity and Hall data (Fig.~\ref{fig:BTSFig1}b).

\subsection{Kramers-Kronig and multi-layer model}
\begin{figure}
\includegraphics[scale=0.95]{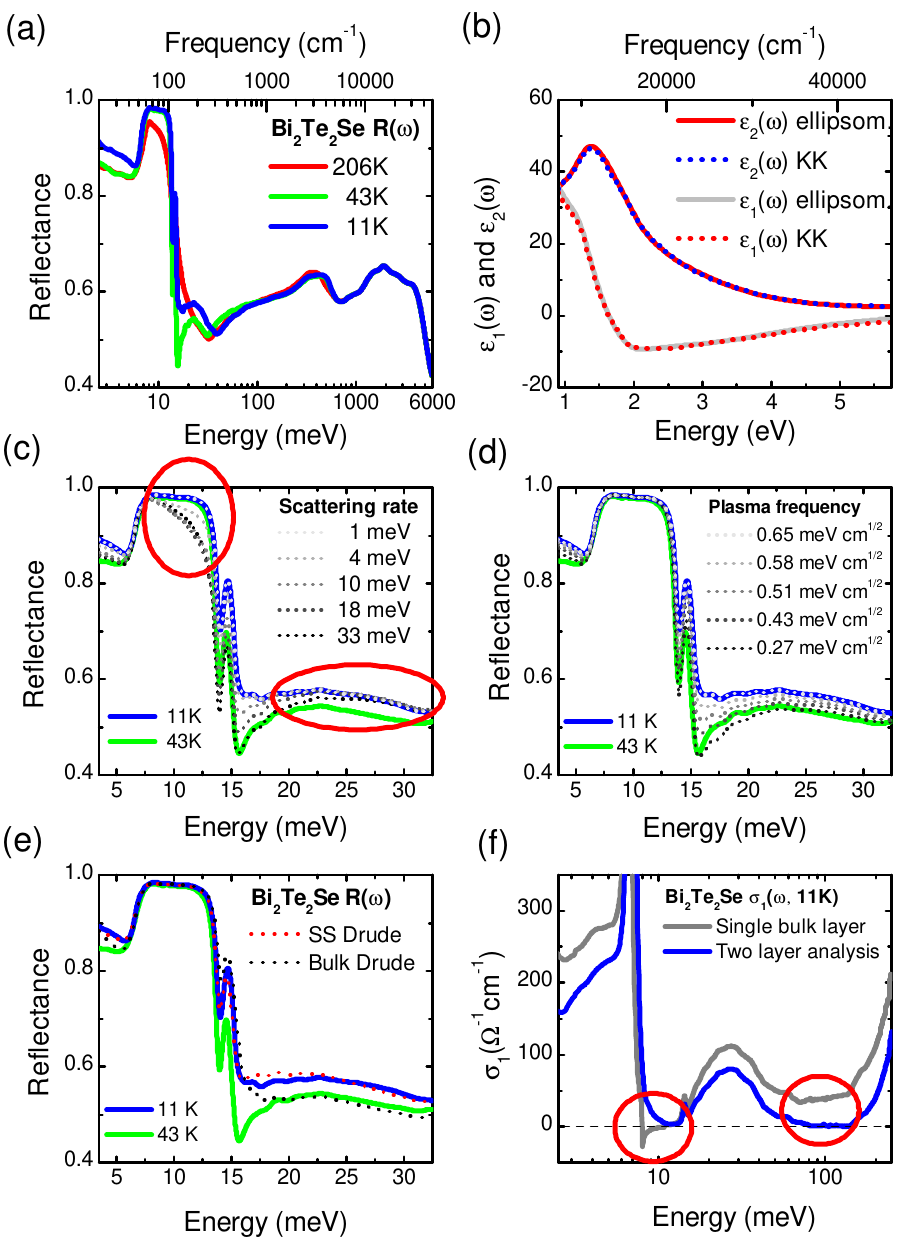}
\caption{\label{fig:BTSFig3}(a) Reflectance of Bi$_2$Te$_2$Se over the full measured range. (b) Comparison of the complex dielectric function obtained by ellipsometry with the complex dielectric function calculated by Kramers-Kronig analysis in RefFit. (c) An increase in surface state Drude scattering rate above 1 meV cannot explain the temperature dependent reflectance spectra. Major discrepancies between the fit and the data are circled in red. (d) A decrease in spectral weight of the surface state Drude fits reflectance data far better. (e) Comparison of the fit quality between a single layer model and a two layer model shows that the increase in reflectance below 43K is much better explained by a Drude mode from the surface. (f) Optical conductivity obtained from single layer analysis (grey) produces unphysical negative optical conductivity, and a sharp artifact between 8-10 meV. Finite conductivity near 100 meV is also unexpected.}
\end{figure}

Optical constants of Bi$_2$Te$_2$Se were obtained from reflectance and ellipsometry data by performing Kramers-Kronig constrained Variational Dielectric Function (VDF) analysis using RefFit software.\cite{KuzmenkoAB:2005jh} Both absolute reflectance and the complex dielectric function obtained by ellipsometry were simultaneously fitted with 2000 oscillators, producing a $\chi^2\ll1$ for each temperature. Comparison between the dielectric function obtained by ellipsometry and the dielectric function found after VDF analysis shows excellent agreement, confirming the accuracy of the VDF analysis (Fig.~\ref{fig:BTSFig3}b). 
To account for the Bi$_2$Te$_2$Se surface states, a multi-layer model was invoked, consisting of a 1 nm thick surface layer\cite{Zhang:2009ks} on top of 0.7 mm bulk Bi$_2$Te$_2$Se (total measured sample thickness). As described later, restricting the model to a single layer resulted in unsuccessful fits and produced unphysical results. We also note that our choice of surface state thickness, while based on theoretical predictions,\cite{Qi:2011wt} had little effect on the results of our analysis. This can be understood by considering the effective conductivity of TIs as described in the Appendix. Finally, a transmission measurement of the Bi$_2$Te$_2$Se crystals revealed complete opacity, rendering the inclusion of the second surface state layer unnecessary. 

Successful two layer fits in $R(\omega)$ were found by adding a Drude term to the otherwise identical surface layer, allowing us to determine the 2D surface plasma frequency $\omega_{p,2D}$ and scattering rate $\gamma$, in addition to the bulk response. As an illustration, Fig.~\ref{fig:BTSFig3}c,d show pre-VDF models that were fitted for 11 K, after which only the surface Drude was altered to capture the temperature dependence of R($\omega$), either by increasing $\gamma$ or decreasing $\omega_{p,2D}$. For the final VDF fit, both bulk and surface parameters were variables. We found that while changes in $\gamma$ below 1 meV could not be discerned from our reflectance data (below our experimental cutoff), an increase above 1 meV could not explain the rise in $R(\omega)$ above 20 meV, and failed to capture the shape of the bulk phonon. However, a decrease in $\omega_{p}$ does successfully model the temperature dependence of the reflectance, as shown in Fig.~\ref{fig:BTSFig3}d. 

Thus, to illustrate the effect of temperature on the kinetic energy of the surface states, without introducing arbitrary fluctuations as a result of local minima in the fitting procedure, the scattering rate was excluded from the fitting parameters, and set to 1 meV (in accord with theoretical predictions\cite{Giraud:2011bc}, and experimental data \cite{2013NatCo...4E2040K,Gehring:2012ue}). We emphasize that this value constitutes an upper bound for $\gamma$, and thus a lower bound for the optical lifetime, since the optical lifetime is less sensitive to small angle scattering compared to the quasiparticle self energy (measured by ARPES and quantum oscillations). \cite{1985PhRvB..32.8442D,1997CaJPh..75..509M} Furthermore, the major contributor to transport lifetime decay (backscattering) is suppressed in TIs. Thus, we expect our optical lifetime to exceed the quasiparticle lifetime. This is confirmed by a comparison of our surface conductance (directly proportional to $\tau$) with previously published values obtained by quantum oscillations (See {\bf Optical Conductivity and Analysis}). Moreover, since the application of a strong magnetic field (required for quantum oscillations) breaks time-reversal symmetry and thus destroys surface state backscattering protection, \cite{1985PhRvB..32.8442D,1997CaJPh..75..509M} comparison of lifetimes/mobilities obtained by different probes is further complicated.

\subsection{Further optical signatures of surface states}
Single-layer modelling was also attempted to explore the possibility of bulk carriers causing the reflectance upturn. The black dotted line in Fig.~\ref{fig:BTSFig3}e shows a single-layer model (before VDF), in which a Drude term is added to the bulk optical properties observed at 43 K, in an attempt to fit the 11 K data. The multi-layer model is shown  in the same figure (red dotted line), comprising a bulk layer with the same optical properties as observed at 43 K, plus a 1 nm surface state layer. The optical properties of the surface layer are identical to the bulk, with an added surface state Drude term. It is clear that the single-layer model fails to capture the FIR increase in $R(\omega)$ below 43 K, even before performing VDF analysis, while the two-layer model produces a superior fit. The inadequacy of the single-layer model becomes even more apparent upon performing the VDF analysis. Fig.~\ref{fig:BTSFig3}f compares $\sigma_1^{Bulk}$ obtained from the VDF analysis for both the single- and two-layer models, revealing negative conductivity and an unusually sharp artifact between 8-10 meV in the former (encircled in red). 
In addition, the single layer model suggests finite conductance near 100 meV, for which we have no explanation. Hence, the unphysical optical conductivity, as well as the poor reflectance fit obtained by single-layer analysis further validate our multi-layer model, revealing surface conductance below 43 K.

Furthermore, we exclude the attribution of the observed conductive channel below 43 K to an accumulation layer on grounds of lacking quantum well state signatures in our ARPES data,\cite{Checkelsky:2009ub,Bianchi:2011kc} the close match between our ARPES and optics plasma frequency (see {\bf Optical Conductivity and Analysis}), and the missing interband transitions that would result from the presence of a quantum well state below the direct gap.\cite{Liu:1993vu}

Finally, to further corroborate our observation of surface conductance in Bi$_2$Te$_2$Se by experimental means, we measured the temperature dependent reflectance spectra of BiSbSe$_2$Te as a counter-example. Our ARPES measurement on this isostructural compound reveals a Fermi energy near the Dirac point, inside the bulk band gap (inset of Fig.~\ref{fig:BTSFig2}d). Hence, surface conductance is expected to be low, (analogous to graphene with $E_F$ at the charge neutrality point\cite{Novoselov:2005es,Li:2008iq}) and likely below our measurement sensitivity due to strong bulk phonons. In addition, transport measurements of BiSbSe$_2$Te reveal even greater longitudinal resistivity at low temperatures compared to Bi$_2$Te$_2$Se, and the Hall coefficient does not flip sign at low temperatures.\cite{Ren:2011vt} Hence, we do not expect to observe a temperature dependent upturn in reflectance, as seen in Bi$_2$Te$_2$Se. Indeed, Fig.~\ref{fig:BTSFig2}d shows that the temperature dependent reflectance spectra of BiSbSe$_2$Te (obtained from the same crystal as the ARPES measurement) are qualitatively similar to Bi$_2$Te$_2$Se, but without a low temperature change in the reflectance minimum or slope.

\subsection{Optical Conductivity and Analysis}
\begin{figure}
\includegraphics[scale=0.94]{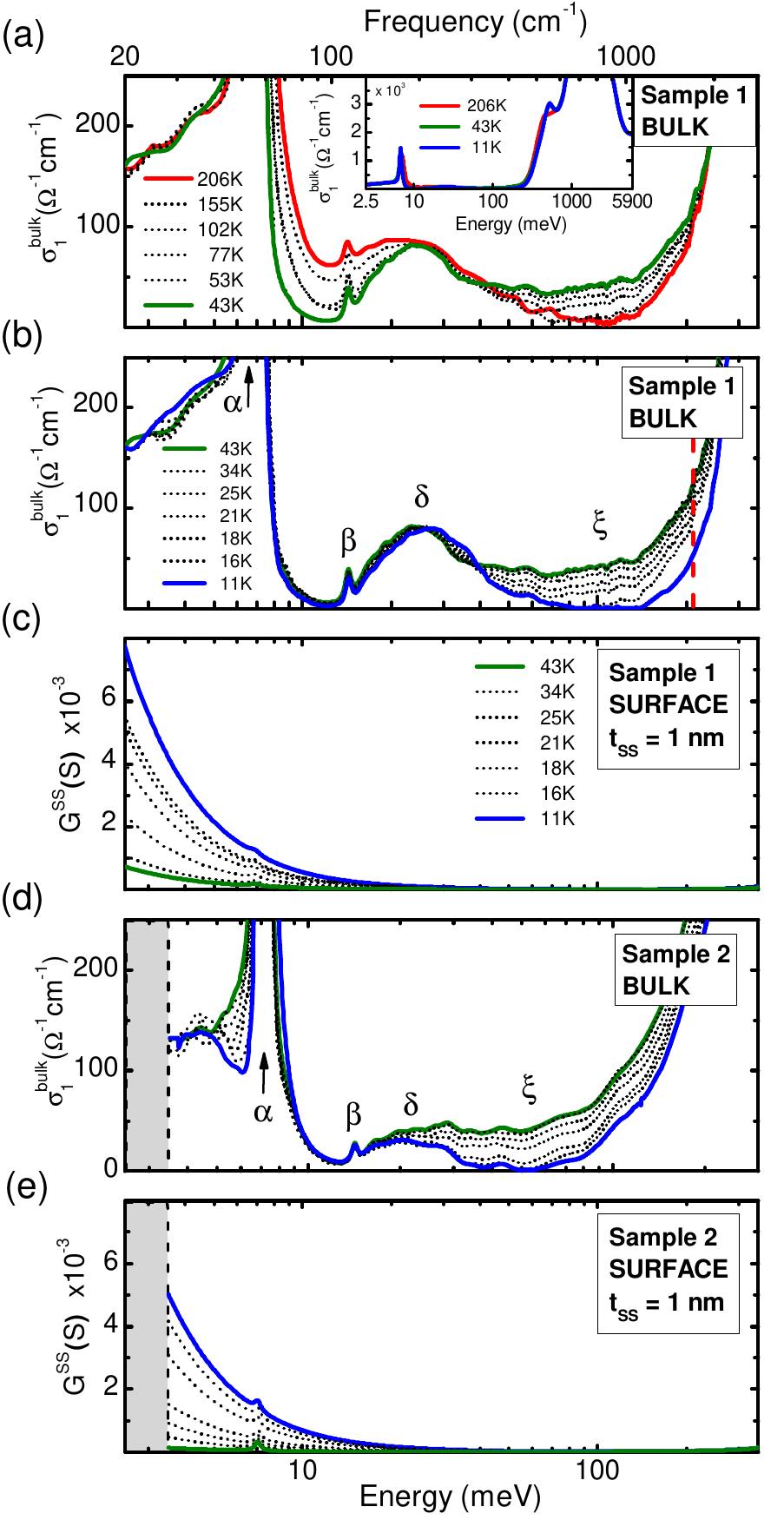}
\caption{\label{fig:BTSFig4}Spectra of the real part of the Bi$_2$Te$_2$Se bulk optical conductivity are shown at high (a) and low (b) temperatures to emphasize their evolution. (c) The real part of the surface conductance of Bi$_2$Te$_2$Se shows the emergence of a Drude mode below 43 K, associated with coherent surface transport. The inset of (a) shows bulk optical conductivity over the full measured range. The real part of the bulk optical conductivity and surface conductance of the second Bi$_2$Te$_2$Se sample are shown in (d) and (e), respectively. }
\end{figure}

For a more quantitative understanding of the optical properties of Bi$_2$Te$_2$Se, we will now discuss its temperature dependent reflectance (Fig.~\ref{fig:BTSFig2}b,c) in conjunction with the real part of the bulk optical conductivity, $\sigma_1(\omega)$ (Fig.~\ref{fig:BTSFig4}a,b,d), and surface conductance, $G^{SS}(\omega)$ (Fig.~\ref{fig:BTSFig4}c,e). Both figures are divided into a high (206 K $\rightarrow$ 43 K) and a low (43 K $\rightarrow$ 11 K) temperature regime for clarity.

{\bf High temperature regime:} All Bi$_2$Te$_2$Se reflectance spectra are dominated by a strong phonon at 7 meV ($\alpha$-mode) and a weaker phonon at 14 meV ($\beta$-mode), both of $E_U$ symmetry.\cite{RICHTER:1977tn} While cooling from 206 K to 43 K, sharpening of the $\alpha$-mode is evident from the change in the slope of the reflectance between 10-30 meV. Fig.~\ref{fig:BTSFig4}a shows direct correlations in $\sigma_1^{bulk}(\omega)$, where the $\alpha$-mode sharpens (asymmetrically) as temperature decreases. This phonon asymmetry suggests quantum interference between the phonon and an electronic continuum, and would thus be best described by a Fano lineshape.\cite{Fano:1961zz} Furthermore, this implies strong electron-phonon coupling, which is somewhat surprising given the low carrier density of this material, but has also been observed in other TIs.\cite{LaForge:2010dx,DiPietro:2012uc} However, we were not able to find a unique fit of the complex optical conductivity between 2.5-10 meV, using a Drude mode (to capture bulk free carriers) and a Fano resonance (to capture the phonon strongly coupled to the carriers).
Thus, the possibility of a feature at finite frequency below that of the phonon, causing the perceived asymmetrical lineshape, cannot be excluded and requires further study.
\

Above 35 meV, an overall increase in reflectance is observed upon cooling to  43 K ($\xi$ feature), which corresponds to increased optical conductivity beyond 40 meV. Since ARPES data shows that the conduction band minimum (CBM) of Bi$_2$Te$_2$Se is at the $\Gamma$-point in the Brillouin zone but the valence band maximum (VBM) is offset from the $\Gamma$-point,\cite{2012PhRvB..85w5406N} we expect to observe indirect gap transitions below the direct band gap energy. Hence, a natural explanation for the $\xi$ feature is an indirect gap transition, which is confirmed by further analysis discussed later. Finally, the $\delta$-mode in R($\omega$), which is revealed by the change in slope between 10-30 meV, is also clearly visible in $\sigma_1^{bulk}(\omega)$, with an onset at 12 meV, and maximum around 25 meV. Similar features have been observed in other TIs,\cite{DiPietro:2012uc,LaForge:2010dx,2012arXiv1209.3593A} and were attributed to an impurity band (IB) in the gap.\cite{DiPietro:2012uc,2012arXiv1209.3593A} However, the temperature independence of the $\delta$-mode is unusual for an IB, as thermal activation from the bulk into the impurity band typically reduces the IB oscillator strength in $\sigma_1^{bulk}(\omega)$, \cite{Liu:1993vu,Burch:2008jy} which we do not observe. Other possible explanations for $\delta$ include a flat-band in the gap, a Holstein polaron, or a two-phonon excitation.\cite{YuCard} Studies beyond the scope of this paper are required to determine its true origin.

{\bf Low temperature regime:} Below 43 K, three distinct changes occur in R($\omega$) with decreasing temperature: {\bf i)} Reflectance below 7 meV increases smoothly (associated with increased conductivity); {\bf ii)} the minimum at 16 meV increases in reflectance (associated with increased conductivity); {\bf iii)} reflectance above 35 meV ($\xi$ feature) decreases (associated with decreased conductivity). We attribute items {\bf i} and {\bf ii} to surface states transport, which is confirmed by our model revealing coherent surface state optical conductance, $G^{SS}(\omega,T)$, below 43 K (Fig.~\ref{fig:BTSFig4}c,e). This increase in conductance below 43 K is corroborated by our DC conductivity data, which deviates from a bulk 3D Variable Range Hopping fit below 40 K (inset of Fig.~\ref{fig:BTSFig1}b). Moreover, the sign change in the Hall coefficient (Fig.~\ref{fig:BTSFig1}b) suggests a shift in dominant carrier mobility and/or scattering rates,\cite{singleton} which has been previously associated with surface conductance.\cite{Ren:2010ji,Luo:2011vf} A quantitative comparison of the surface conductance (Fig.~\ref{fig:BTSFig4}c) at 2.5 meV and 11 K ($\approx 8$ mS) with a previous THz study of Bi$_2$Se$_3$ thin films where $G^{SS}($2.5 meV, 10 K) $\approx 7$ mS shows good agreement.\cite{2012PhRvL.108h7403V} Surface conductance obtained using quantum oscillations (up to 4.4 mS), shows lower values, as expected from the fundamentally different particle lifetimes obtained by quantum oscillation experiments.\cite{Hor:2009hl,2010Sci...329..821Q,Xiong:2011tm} We also compare our results with transport and ARPES data as follows.

The Boltzmann transport theory for spin-polarized Dirac bands predicts that the 2D surface conductance is given by\cite{2007PNAS..10418392A}
\begin{equation}\label{SigBOLTZ}
G^{SS}=\left({\frac{e^2}{h}}\right){\frac{E_{F}}{2\hbar\gamma}}g_{s}g_{v}
\end{equation}
where $g_{s}$ and $g_{v}$ are the spin and valley degeneracy, respectively, both equal to 1 for TI surface states. Substituting the Fermi velocity and wave vector for $E_{F}=\hbar v_{F}k_{F}$, and using the standard Drude approach $G^{SS}=\omega_{p,2D}^{2}/4\pi\gamma$, we calculate the 2D plasma frequency using
\begin{equation}\label{WpBOLTZ}
\omega_{p,2D}=\sqrt{\left({\frac{e^2}{h}}\right)2\pi v_Fk_F}
\end{equation}
Indeed, $G^{SS}(\omega,T)$ contains a Drude with $\hbar\omega_{p,2D}=0.64$ meV cm$^{1/2}$ at 11 K. From our low temperature ARPES data, with $k_F=0.03$ $\AA^{-1}$, $v_F=6\times10^{5}$ m/s, we find $\hbar\omega_{p,2D}=0.13$ meV cm$^{1/2}$. Using values from previously published quantum oscillation and ARPES measurements,\cite{Ren:2010ji,2011arXiv1101.1315X} we find $0.14$ meV cm$^{1/2}$ $<\hbar\omega_{p,2D} < 0.17$ meV cm$^{1/2}$. The observed discrepancy likely arises due to multiple surface states being probed as a result of the cleaving process. This has previously been reported in Bi$_2$Te$_2$Se\cite{2012arXiv1209.3593A}, as well as Bi$_2$Te$_3$\cite{Jenkins:2010kg} and will be discussed in detail in a future publication.

As expected from the temperature dependence of the reflectance minimum, and the upturn below 7 meV, we find that coherent free carrier transport at the surface is suppressed as the temperature is raised (Fig.~\ref{fig:BTSFig4}c,e). A four-fold reduction in $\hbar\omega_{p,2D}$ is observed as temperature is increased from 11 K to 43 K. Above 43 K, changes in R$(\omega)$ originating from the surface conductance are concealed by the prominent $\alpha$ phonon. This reduction was confirmed by repeated far infrared measurements, using freshly cleaved surfaces of two different Bi$_2$Te$_2$Se crystals (Fig.~\ref{fig:BTSFig2}b,c). Considering the great resemblance in optical properties of both crystals, we continue our analysis focussing on sample 1, which was larger, and thus could be measured over a greater energy range. 

Interestingly, as the surface conductance is suppressed with temperature, the feature associated with the indirect gap ($\xi$) emerges at the same rate. This can be seen from the enhancement of reflectance above 35 meV (Fig.~\ref{fig:BTSFig2}b,c) as well as the increased spectral weight in $\sigma_1^{bulk}(\omega)$ for $\hbar\omega > 40$ meV (Fig.~\ref{fig:BTSFig4}b,d). 
As discussed later, the onset of this indirect gap transition as a function of temperature is not surprising as this transition requires thermally activated phonons. However, the correspondence between the temperature dependence of the surface conductance and the indirect gap suggest phonons are also responsible for suppressing the surface state transport.

\begin{figure}
\includegraphics[scale=0.94]{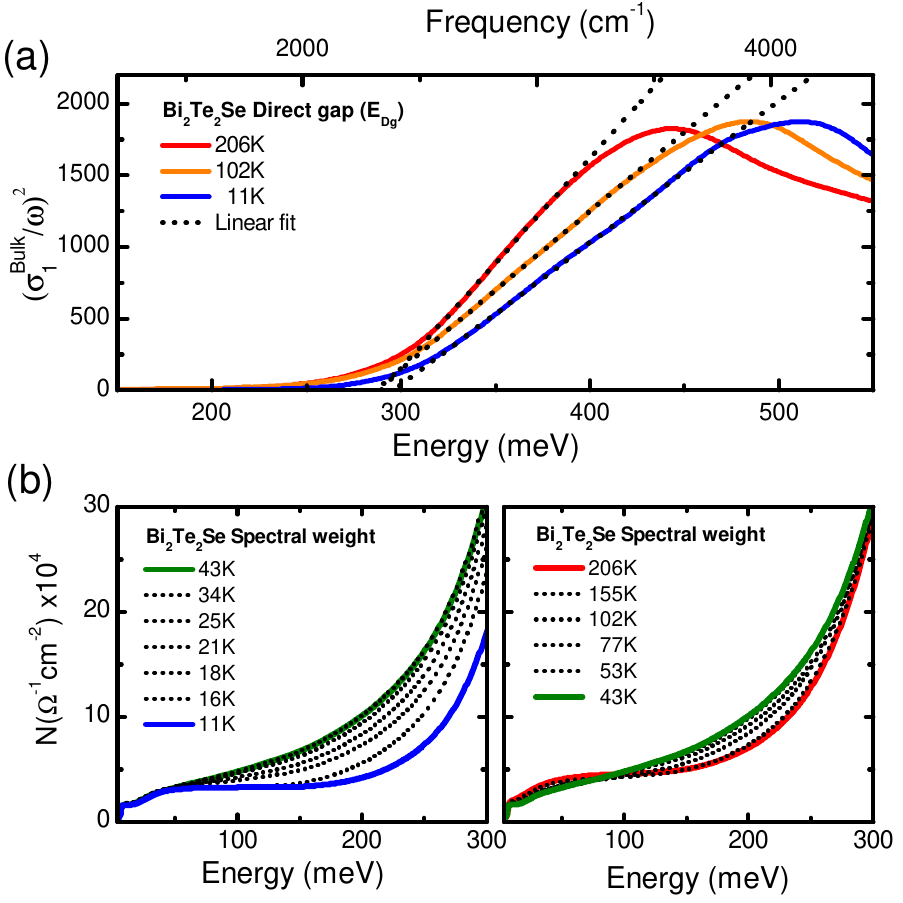}
\caption{\label{fig:BTSFig5}(a) The x-intercept of a linear fit in $(\sigma_1^{bulk}/\omega)^2$ reveals the direct gap transition, E$_{Dg}$, of Bi$_2$Te$_2$Se around 290 meV. (b) Spectral weight of the bulk optical conductivity is shown for spectra between 11 K - 43 K, and 43 K - 206 K. The strong temperature dependence around 210 meV corresponds to an indirect gap transition, E$_{IDg}$.}
\end{figure}

For a better understanding of the temperature dependence seen in Bi$_2$Te$_2$Se it is instructive to examine its band structure. This can be evaluated directly in our optical experiments by estimating the direct bulk band gap ($E_{Dg}$) from a linear fit in $(\sigma_1^{bulk}/\omega)^2$, where ($\sigma_1^{bulk}/\omega)^2\approx C(\omega-E_{Dg})$ in the vicinity of the gap, and $C$ is determined by matrix elements.\cite{YuCard} We find $E_{Dg}$ has a temperature independent value of 290$\pm$7 meV (Fig.~\ref{fig:BTSFig5}a), consistent with recent measurements via ARPES, \cite{2012PhRvL.109p6802M,2012PhRvB..85w5406N} and optics experiments.\cite{Greenaway:1965vl,2012arXiv1209.3593A,DiPietro:2012uc} In contrast with the binary TI Bi$_2$Se$_3$,\cite{LaForge:2010dx,Post:2013td} the absence of a Moss-Burstein shift in $E_{Dg}$ is as expected, considering the in-gap Fermi level of Bi$_2$Te$_2$Se. Since the $\xi$ feature is observed below $E_{Dg}$, it is likely the result of an indirect gap transition.

For further investigation, the spectral weight of $\sigma_1^{bulk}(\omega)$ is calculated (Fig.~\ref{fig:BTSFig5}b) using 
\begin{equation}\label{SW}
N(\omega,T)=\int_{\omega_0}^{\omega_{f}} \! \sigma_1(\omega ',T) \,  \mathrm{d}\omega '
\end{equation}
where $\hbar\omega_0$ is our lowest measured energy of 2.5 meV, and $\hbar\omega_f=210$ meV captures the $\xi$ feature's spectral weight, but is below $E_{DG}$. Fig.~\ref{fig:BTSFig6}d shows the absolute spectral weight reaching a maximum amplitude at 43 K. Neither this temperature dependence, nor the energy scale (210 meV $\gg k_{B}T$) can be explained by an impurity band model, in which spectral weight is expected to monotonically decrease as temperature rises.\cite{Liu:1993vu,2011PhRvB..84h1203C} Hence, a phonon-assisted indirect gap transition ($E_{IDg}$) is a more likely explanation. Broadening of the onset of this indirect transition is likely due to multiphonon absorption or disorder, as seen in Fig.~\ref{fig:BTSFig4}a,b.

\section{Discussion}
\begin{figure}
\includegraphics[scale=1]{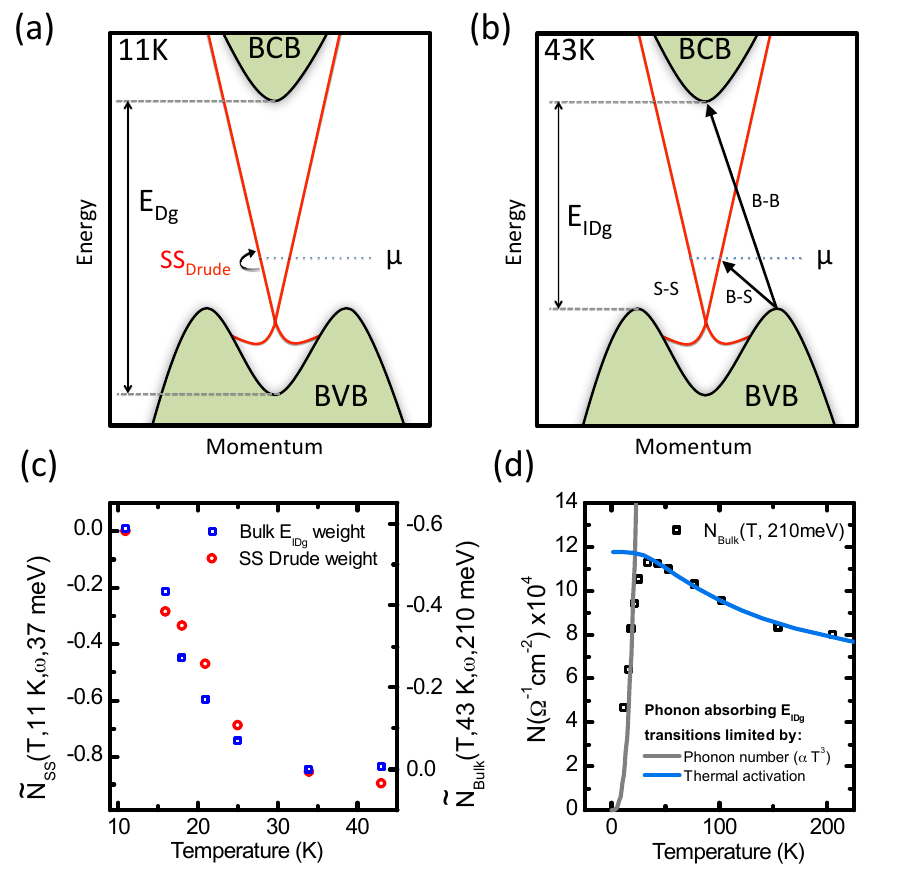}
\caption{\label{fig:BTSFig6}Bandstructure cartoon for Bi$_2$Te$_2$Se at 11 K (a) and 43 K (b). While coherent surface state transport is observed at 11K, phonon-mediated coupling between bulk and surface states suppresses surface transport at 43 K. (c) Comparison of the SS Drude spectral weight and bulk indirect gap (E$_{IDg}$) spectral weight, relative to their respective maxima, reveals a similar temperature dependence. (d) The indirect bandgap's spectral weight at 210 meV shows the competition between two limitations: available phonons below 43 K, and thermal excitation above 43 K.}
\end{figure}

A comprehensive picture of the electronic structure and carrier dynamics of Bi$_2$Te$_2$Se unfolds from our results.
At low temperatures (Fig.~\ref{fig:BTSFig6}a), the chemical potential is located above the bulk valence band, allowing for coherent surface conductance (SS$_{Drude}$), and giving rise to the electron-like Hall signal and metallic behavior in the DC resistivity. In addition, direct excitations from the bulk valence band to the conduction band are observed for $\hbar\omega>E_{Dg}$, while the absence of thermally excited phonons suppresses all phonon-mediated indirect transitions for   $\hbar\omega<E_{Dg}$. As temperature rises to 43 K (Fig.~\ref{fig:BTSFig6}b), the thermal excitation of phonons facilitates two indirect (finite {\bf q}) gap excitations: valence band maximum (VBM) to conduction band minimum (CBM), indicated by B-B in Fig.~\ref{fig:BTSFig6}b, and VBM to surface states, indicated by B-S in Fig.~\ref{fig:BTSFig6}b. 

Since both the B-B and B-S transitions are phonon-mediated, and require similar {\bf q}, the same phonons are likely involved in both transitions. Hence, a strong correlation between their temperature dependences is expected, which should be evident in a spectral weight comparison. However, because the B-S transition is too weak to observe, we use the temperature dependence of the surface conductance as a measure of the B-S transition. Hence, we superimpose the surface conductance spectral weight (relative to 11 K) and the B-B indirect gap transition spectral weight (relative to 43 K) in Fig.~\ref{fig:BTSFig6}c. Here, $\widetilde{N}$ is given by $\widetilde{N}(T,T_{m},\omega_{m})=[N(T,\omega_{m})-N(T_{m},\omega_{m})]/N(T_{m},\omega_{m})$, where $T_m$ is the temperature at which the spectral weight peaks, and $\omega_m$ is the upper limit of spectral weight integration. Indeed, similar temperature dependence is evident from the plot, suggesting phonons cause the suppression of coherent surface conductance. To understand how this works, both phonon-mediated indirect transitions, B-B and B-S, will be discussed separately.

{\bf B-B:} By providing the required momentum ({\bf q}), thermally excited phonons mediate the indirect transition from VBM to CBM, which is clear from the strongly temperature dependent spectral weight gain in Fig.~\ref{fig:BTSFig6}d.
The low temperature fit (grey line) conveys the number of available phonons below the Debye temperature ($N_{ph}\propto T^3$),\cite{singleton} which extrapolates to a spectral weight of 0 $\Omega^{-1}cm^{-2}$ at 0 K. This fit confirms that thermally activated phonons constitute the low temperature limitation of the indirect B-B transition. Minor deviations from the fit are expected as a result of constant spectral weight contributions from features below the indirect gap.

{\bf B-S:} As temperature rises, phonons mediate bulk to surface state coupling, and suppress coherent surface state transport (Fig.~\ref{fig:BTSFig4}c,e). This coupling can be effectively understood as phonons providing the necessary momentum to excite electrons from the valence band into (B$\rightarrow$S) and out of (S$\rightarrow$B) the surface states in a continuous cycle. Consequently, surface carriers lose kinetic energy and the surface Drude spectral weight is transferred to higher energies as total spectral weight must be conserved.\cite{2007NJPh....9..229K} We emphasize, however, that we see no evidence for an increase in the surface scattering rate (i.e. a broadening surface Drude), though this may occur below our resolution, or low-energy cut-off in R($\omega$). Hence, we conclude that spectral weight must be transferred to (in)direct surface state transitions at finite energy, with a width and amplitude that lies below our detection sensitivity.\cite{2007NJPh....9..229K,1999Sci...283...49B,2004Sci...304..708B} This can be understood by replacing a Lorentzian centred at 0 meV with a Lorentzian centred at finite frequency, and changing its 1 meV width (upper bound for $\gamma$) to a width of 0.5 eV (reasonable for a narrow interband transition), while keeping $\omega_p^2$ constant. This $\approx$ 500-fold reduction in $G^{SS}$, especially with strong bulk conductivity over the same energy range, is even less visible in $R(\omega)$, and thus below our detection limit. Nevertheless, the reduced reflectance below the phonon as temperature rises (Fig.~\ref{fig:BTSFig2}b,c), is a clear indication of reduced surface conductance. It is worth noting that this effect does not constitute a temperature dependent transmutation from a topologically non-trivial to trivial band insulator, as was recently predicted to occur under special circumstances.\cite{PRL:Garate2013} This would violate the robust time reversal symmetry protection of TI surface states (no magnetic field is applied), and would have been observed in our optics measurements as we directly probe the bulk band gap (and see no evidence of band inversion). 

Above 43 K, we observe a decline in the B-B indirect gap transition spectral weight. This is expected since thermal activation of the bulk causes the depletion of electrons at the VBM, thus reducing the number of available electrons for this transition. A fit of $N$(210 meV,T) above 43 K (blue line in Fig.~\ref{fig:BTSFig6}d), shows that this decline obeys the Arrhenius law, with an activation energy of 15 meV. This value matches the activation energy of 16 meV, obtained from an Arrhenius fit of the DC resistivity (black dotted line in Fig.~\ref{fig:BTSFig1}b), as expected.
It should be noted that thermal activation of the VBM is an intricate function of not only temperature, but also the bulk and surface density of states, total carrier density, and the Fermi-Dirac distribution. Minor variations among Bi$_2$Te$_2$Se crystals can thus account for the spread in observed activation energies.\cite{Sobota:2012bt,Luo:2011vf} Hence, while the Arrhenius law captures the trend of both $N$(210 meV,T) and $\rho_{xx}$(T) nicely, its fitting parameters are not conclusive. Further study is required to reveal the details of the thermal activation. 

Since thermal activation of the bulk causes the depletion of electrons at the VBM, it also results in bulk hole transport. We therefore conclude that while metallic surface state electron transport persists up to 43 K, phonon-mediated thermal activation of the VBM causes bulk hole-like transport to dominate above 43 K. This is confirmed by the change in Hall sign, followed by a peak and subsequent decline in $R_{H}(T)\sim 1/n$, indicative of an increasing hole density (Fig.~\ref{fig:BTSFig1}b). Moreover, above 43 K the temperature dependent DC conductivity (inset of Fig.~\ref{fig:BTSFig1}b) clearly shows VRH behavior, associated with bulk states,\cite{Ren:2010ji} and we see no evidence of a surface state Drude.

Finally, while phonon induced surface to bulk coupling is also likely to affect the quasiparticle lifetime, further studies probing below our low energy cut-off are required to quantify this effect. Theoretical work by Giraud and Egger predicts that the surface quasiparticle decay rate increases by more than an order of magnitude between 10 K and 50 K due to surface state scattering off acoustical phonons.\cite{Giraud:2011bc} In addition, optical spectroscopy of several other TIs, in which phonons are best described by a Fano lineshape, as well as transport measurements, indicate that electron-phonon coupling in TIs is strong.\cite{LaForge:2010dx,DiPietro:2012uc,Kim:2012cs} Interestingly, ARPES results are still controversial. While several studies find evidence of electron-phonon coupling in TIs,\cite{Kondo:2013vw,Sobota:2012bt,2013NatSR...3E2411C,Howard:2013uz} one group reached different conclusions by studying the temperature dependent surface state self-energy.\cite{2012PhRvL.108r7001P} This is perhaps not surprising since we find a surface scattering rate upper bound of 1 meV, which is below the resolution of the described ARPES experiments. Another explanation can be found in the more subtle difference between the ARPES measured quasiparticle lifetime and the scattering lifetime obtained by optics, as described earlier in section {\bf Kramers-Kronig and multi-layer model}. \cite{1997CaJPh..75..509M,1985PhRvB..32.8442D} Since the quasiparticle lifetime underestimates the optical scattering lifetime,\cite{Ren:2010ji} direct comparison is challenging. This suggests further work is required to extract the true lifetimes of surface states in both ARPES and optics. Furthermore, the phonon mediated surface to bulk coupling is very sensitive to the position of the chemical potential and the details of the electron and phonon dispersions in a given compound, as recently demonstrated via ARPES.\cite{2013NatSR...3E2411C} Specifically, if the Fermi energy is sufficiently far away from the VBM, thermal activation of bulk states into the surface would require higher temperatures, and surface to bulk coupling could only occur through multiple phonon absorption.
\

\section{Conclusion}
In conclusion, we have performed temperature dependent broadband optical spectroscopy measurements of four Bi based topological insulators, and focussed specifically on Bi$_2$Te$_2$Se. At low temperatures, we find strong optical evidence of surface state conductance (i.e. a characteristic upturn in the FIR reflectance below 43 K, and a reflectance minimum increase at 16 meV), corroborated by a unique multi-layer physical fit of our Reflectance data. Quantitative comparison of the obtained surface conductance with previously published optics data shows good agreement, while the discrepancy with the Boltzman transport prediction is likely due to the inadvertent measurement of multiple surface resulting from the cleaving proces.\cite{Ren:2010ji,Luo:2011vf,2012PhRvL.108h7403V,2007PhRvL..98r6806H,2011arXiv1101.1315X,Ren:2011vt,2012PhRvL.109p6802M,2012PhRvB..85w5406N,Greenaway:1965vl,2012arXiv1209.3593A,DiPietro:2012uc,Sobota:2012bt,Giraud:2011bc,Jenkins:2010kg} Further confirmation of our optical data comes from DC transport measurements on the same Bi$_2$Te$_2$Se crystal, where we find a distinct metallic conduction channel below 43 K in our longitudinal DC conductance data. Finally, we corroborate our observation of surface states by a counterexample, measuring the optical and ARPES response of isostructural compound BiSbSe$_2$Te. In this material, where the Fermi level coincides with the (exposed) Dirac point, we expect no optical signatures of surfaces states (analogous to graphene measurements with E$_F$ at the charge neutrality point\cite{Novoselov:2005es,Li:2008iq}), which is confirmed by our temperature dependent reflectance data. 

Upon warming, we find that surface state kinetic energy is suppressed (i.e. surface free carrier spectral weight decreases), and  observe a concurrent sign change in the Hall coefficient, suggesting a switch from surface electron dominated transport to bulk valence band hole transport. This is confirmed by VRH behavior in the DC conductivity above 43 K, associated with bulk carriers.\cite{Ren:2010ji} Moreover, spectral weight analysis shows that the rate of surface kinetic energy suppression matches the rate at which the indirect bulk gap spectral weight grows. Hence, since indirect bulk gap transition are strictly possible in the presence of phonons, and the indirect bulk gap spectral weight follows $T^3$ (available number of phonons below $T_{Debye}$), our results suggest that electron-phonon coupling plays a central role in the surface state suppression.

Our findings also illustrate a significant reduction of bulk conductivity at low temperatures in Bi$_2$Te$_2$Se, and BiSbSe$_2$Te, compared to binary compounds Bi$_2$Se$_3$ and Bi$_2$Te$_3$. With Bi$_2$Te$_2$Se showing the most prominent signatures of low temperature surface conductance, we find this TI to be the strongest candidate for further study of the electronic properties of TI surface states. Nevertheless, while these novel (qua-)ternary compounds are an exciting step closer towards the isolation of surface state dominated transport for device applications, our results emphasize the significance of electron-phonon interactions in TIs as a non-trivial limitation to room temperature operation. Our work clearly shows that the suppression of electron-phonon coupling along with a reduction in bulk carriers of future TIs are required for successful devices and the fundamental study of surface state transport properties.

\

\section{Acknowledgements}
We acknowledge D. Basov, and N. Nagaosa, for very helpful discussions, and Young-June Kim for use of an ellipsometer. The work at the University of Toronto was partially funded by the Ontario Research Fund, the Natural Sciences and Engineering Research council of Canada, Canada Foundation for Innovation, and the Prins Bernhard Cultuurfonds. The work at Princeton by SJ, MEC and RJC was supported by SPAWAR grant N6601-11-1-4110. The Princeton-led synchrotron X-ray-based ARPES measurements (SX, MN, MZH) are supported by the Office of Basic Energy Sciences, U.S. Department of Energy (grants DE-FG-02-05ER46200, and AC03-76SF00098). XW and SWC at Rutgers University are supported by the NSF under Grant No. NSF-DMR-1104484. 

%

\newpage
\appendix

\begin{figure}
\includegraphics[scale=0.15]{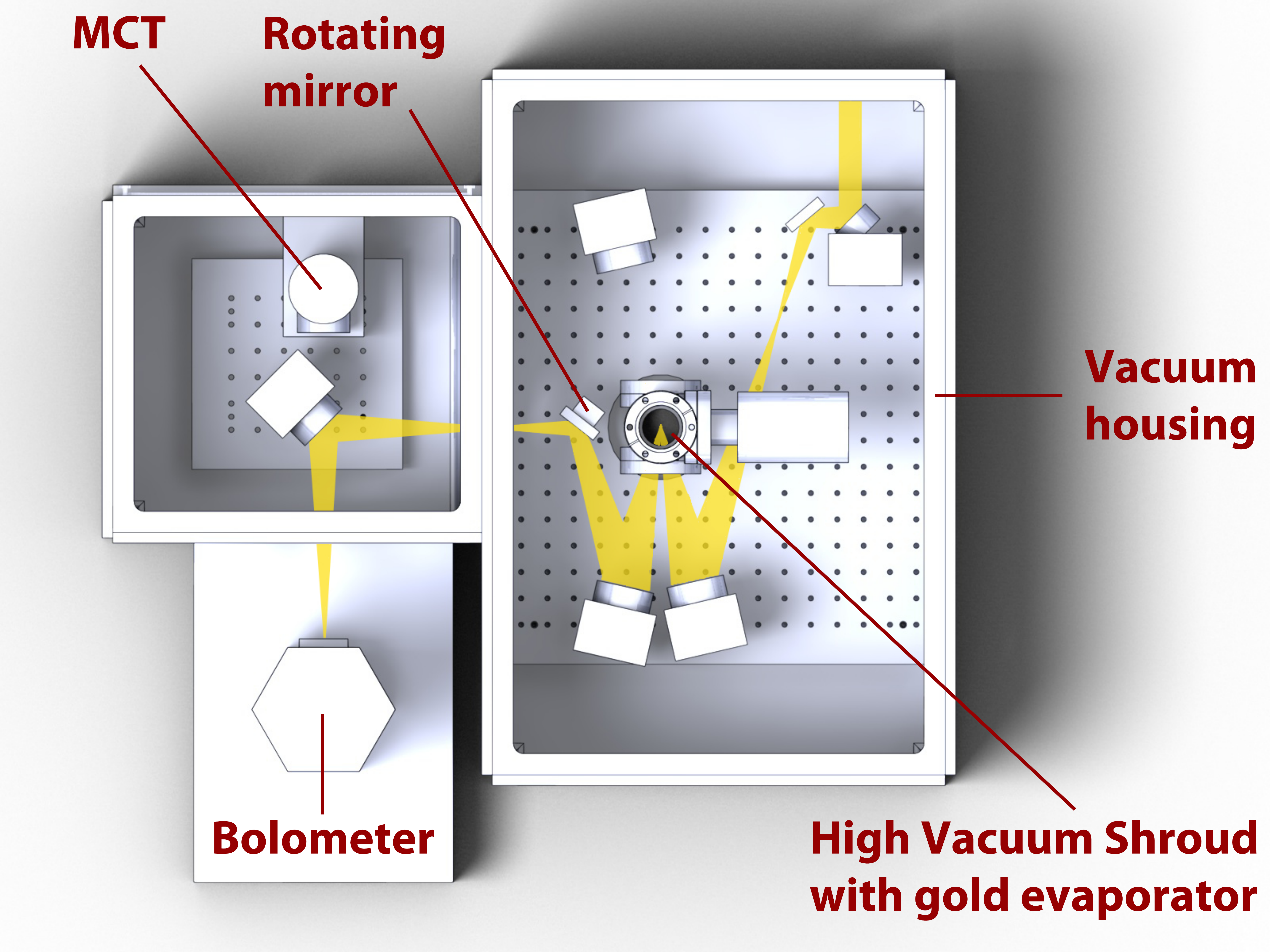}
\caption{\label{fig:zerofieldref}IR beam path of the custom designed extension to a Bruker VERTEX 80v spectrometer. A thermal evaporator is attached to the shroud for in-situ gold coating, and the rotating mirror facilitates near-simultaneous measurement of reflectance and transmittance. Mirrors and optical layout were designed with Advanced Systems Analysis Program ray-tracing software.}
\end{figure}

\section{Experimental setup}

All reflectance measurements were performed in near-normal geometry over the far through near-infrared range (2.5 meV - 1.24 eV), using a modified Bruker VERTEX 80v FTIR Spectrometer. Fig.~\ref{fig:zerofieldref} shows the IR beam path of the homebuilt extension coupled to the spectrometer.  To produce absolute reflectance spectra and correct for unintentional variances between sample and reference (e.g. surface roughness, size, alignment, etc.),  the high vacuum sample chamber is equipped with a thermal evaporator for in-situ gold coating.

While this report only discusses bulk reflectance measurements, the rotating mirror in the extension allows for nearly simultaneous reflectance and transmittance measurements. Additionally, the extension is equipped with polarizers, a 405 nm, 200 mW CW laser, a chopper, and 8 vacuum feedthroughs to the sample space, making this setup highly versatile. 

\begin{figure}
\includegraphics[scale=0.19]{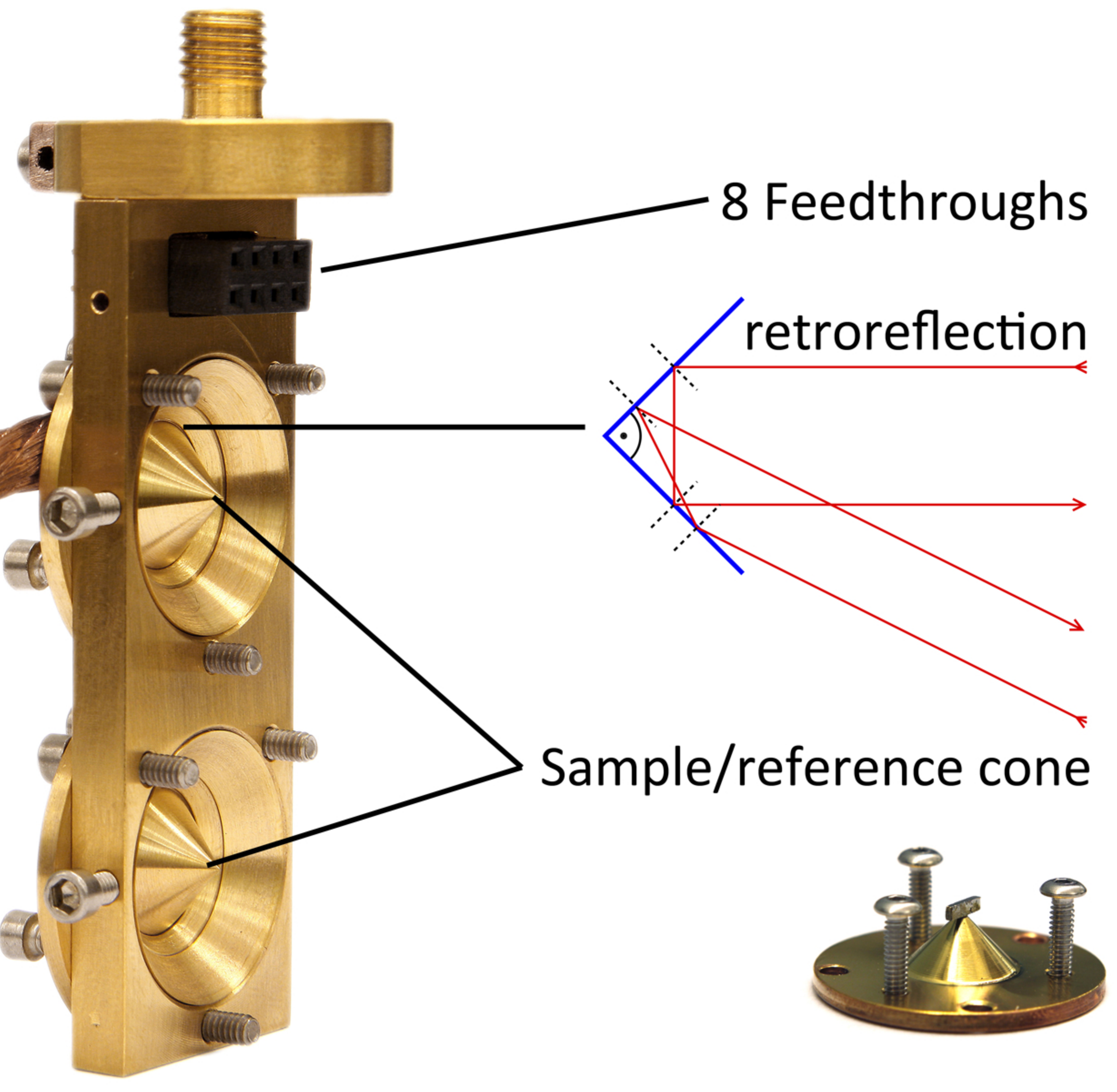}
\caption{\label{fig:ColdfingerCone}Left, a picture of our coldfinger enabling both reflectance and transmittance measurements, configured with a cone for reflectance. Top right, a diagram indicating the retroreflection preventing spurious signals from reaching the detector. Bottom right, a Bi$_2$Se$_3$ sample mounted on the cone for reflectance measurements along the c-axis.}
\end{figure}

Each sample was mounted on a gold coated cone with thermally conductive STYCAST 2850 FT. Both reference and sample cones were then attached to a custom made coldfinger (Fig.~\ref{fig:ColdfingerCone}), designed to reduce the detection of unwanted radiation. The 90 degree angle between the cone and coldfinger ensures that radiation that misses the sample is specularly retro-reflected.  A copper braid thermally connects the back of the cone with the body of the coldfinger, and a Cernox temperature sensor on the cone monitors the temperature of the sample. The sample and reference are laser aligned with respect to the coldfinger (and each other) before entering the shroud.

\begin{figure}
\includegraphics[scale=0.95]{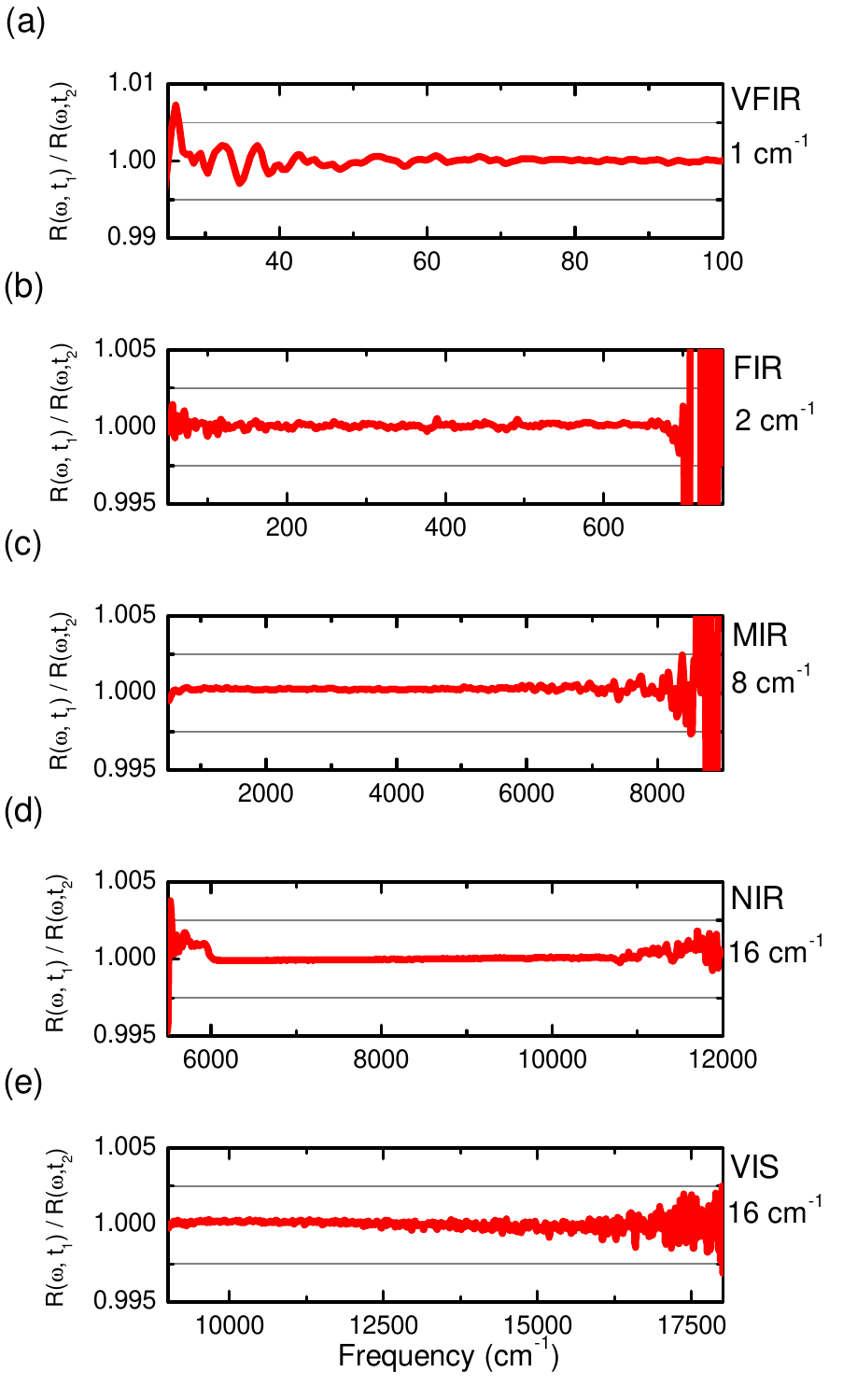}
\caption{\label{fig:100percentzerofield}Characterization of custom FTIR setup. (a), (b), (c), (d), (e), 100\% percent lines over the full frequency range of our FTIR system, showing excellent reproducibility. The resolution of each frequency range is listed to the right of the figures.}
\end{figure}

\begin{figure}
\includegraphics[scale=0.89]{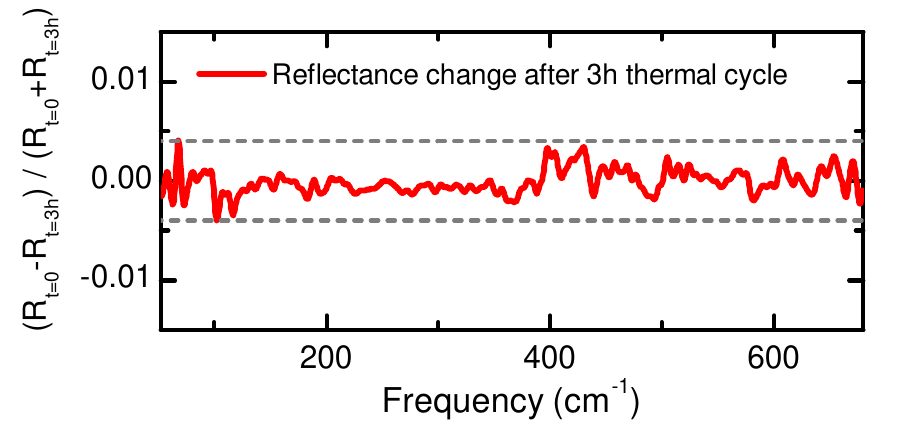}
\caption{\label{fig:BSrepeat}FIR Reflectance reproducibility of a 1x4 mm mechanically polished ac-face of Bi$_2$Se$_3$ is better than 0.4\%, even after a period of 3 hours during which temperature was cycled from 30 K to 292 K, and back.}
\end{figure}

For the observation of subtle changes in temperature dependent reflectance, stability of the setup is pivotal. Fig.~\ref{fig:100percentzerofield}(a)-(e) shows 100\% lines (R($\omega,t_1$)/R($\omega,t_2$), where $t_2-t_1\approx 60$ $s$) of our setup over the full accessible spectral range. Each range is indicated by its abbreviations, and the resolution of the measurement is listed to the right of each spectrum (i.e. the Very Far InfraRed (VFIR) was measured at a resolution of 1 cm$^{-1}$, Far InfraRed (FIR) at 2 cm$^{-1}$, etc.). Even after several hours and a thermal cycle from 30K to 292K and back, FIR reproducibility is still better than 0.4\% (Fig.~\ref{fig:BSrepeat}). 

To supplement the reflectance spectral range and minimize extrapolation errors, a Woollam Inc VB-400 VASE ellipsometer was used between 0.75-5.95 eV. Ellipsometry provides direct access to the complex dielectric function, which was fitted simultaneously with the reflectance data during the RefFit analysis. This simultaneous fitting technique minimizes errors in the high frequency reflectance extrapolation, compared to using only reflectance.

\newpage

\section{Effective optical conductivity}
Using the Drude model to capture the surface conductance, successful two-layer fits in R($\omega$) could be obtained over a range of surface state thicknesses $t_{SS}$, as long as $t_{SS}$ remained much smaller than the penetration depth of light $\delta(\omega)$, and the 2D surface state plasma frequency was kept constant. This can be understood by approximating the effective optical conductivity as
\begin{equation}\label{EffMed}
\hat{\sigma}_{eff}=f\hat{\sigma}_{Bulk}+(1-f)\hat{\sigma}_{SS}
\end{equation}
where $f$ is the probed volume fraction of $\hat{\sigma}_{eff}$ attributabed to bulk states ($\hat{\sigma}_{Bulk}$), and a $(1-f)$ fraction is attributed to surface states ($\hat{\sigma}_{SS}$). Since $f$ is determined primarily by $t_{SS}$ and $\delta(\omega)$ we can approximate $(1-f)\approx t_{SS}/\delta(\omega)$. As long as $\omega\gg\gamma$ we have $\hat{\sigma}=i\omega^{2}_{p,3D}/\omega$, where $\omega^{2}_{p,3D}$ is the 3D plasma frequency. The 2D surface state plasma frequency is given by $\omega^{2}_{p,2D}=\omega^{2}_{p,3D}*t_{SS}$, and thus we find $(1-f)\hat{\sigma}_{SS}=i\omega^{2}_{p,2D}/\omega\delta(\omega)$. This implies that as long as $t_{SS}\ll\delta(\omega)$, we can approximate the optical conductivity as
\begin{equation}\label{EffMed2}
\hat{\sigma}_{eff}\approx f\hat{\sigma}_{Bulk}+\frac{i\omega^{2}_{p,2D}}{\omega\delta(\omega)}
\end{equation}
In other words, the measured surface state plasma frequency will be independent of the actual thickness of the surface states. Hence, to obtain the surface state conductance we assume $t_{SS}=1$ nm based on theoretical predictions,\cite{Qi:2011wt} though this assumption has little effect on the results of our analysis.

\end{document}